\documentclass[aps,prl,twocolumn,groupedaddress,showpacs,tighten]{revtex4}
\usepackage{graphicx}
\usepackage{times}

\def\be{\begin{equation}}
\def\ee{\end{equation}}
\def\bea{\begin{eqnarray}}
\def\eea{\end{eqnarray}}
\def\ba{\begin{array}}
\def\ea{\end{array}}
\def\ghost#1{}

\def\beq{\begin{equation}}
\def\eeq{\end{equation}}
\def\bey{\begin{eqnarray}}
\def\eey{\end{eqnarray}}

\def\lsim{\mathrel{\raise.3ex\hbox{$<$\kern-.75em\lower1ex\hbox{$\sim$}}}}
\def\gsim{\mathrel{\raise.3ex\hbox{$>$\kern-.75em\lower1ex\hbox{$\sim$}}}}

\begin{document}

\title{\boldmath $U$-boson detectability, \,and Light Dark Matter}
\author{Pierre Fayet$^{1}$}

\affiliation{$^1$Laboratoire de Physique Th\'eorique de l'ENS, UMR 8549 CNRS, 24 rue Lhomond, 75231 Paris Cedex 05, France}
\date{\today}

\begin{abstract}
{
The possible existence of a new gauge boson $U$, light and very weakly coupled, 
allows for Light Dark Matter particles, which could also be at the origin of the 511 keV line 
from the galactic bulge. Independently of dark matter, and taking into account possible $Z$-$U$ mixing effects, we show that, even under favorable circumstances
(no axial couplings leading to an axionlike behavior or extra parity-violation effects,
very small coupling to $\nu$'s),
and using reasonable assumptions (no cancellation effect in $g_\mu-2$, 
lepton universality), 
\,the $\,U$ coupling to electrons can be at most as large as $ \simeq $ $1.5\ 10^{-3}$ 
\,(for $m_U<m_\mu$),
and is likely to be smaller (e.g. $\lsim 3\,10^{-6}$ $m_U$(MeV), 
if the $U$ couplings to $\nu$ and $e$ are similar).
This restricts significantly the detectability of a light $U$ in $\,e^+e^-\to\gamma\,U$,
in particular.
$\,U$ exchanges can still provide annihilation cross sections of LDM particles 
of the appropriate size, even if this may require that light dark matter 
be relatively strongly self-interacting.
}
\end{abstract}
\pacs{\ 12.60.Cn, 13.66.Hk, 14.70.Pw, 
95.35.+d, 95.85.Pw
\hspace{20mm} LPTENS-06/24}
\maketitle

\vspace{-5mm}
Theories beyond the Standard Model often involve extended gauge groups, 
neces\-sitating new spin-1 gauge bosons,
in addition to the gluons, photon, $W^\pm$ and $Z$.
It is usually believed that they
should be heavy 
($\gsim $ several hundred GeV's at least) or even very heavy, as in grand-unified theories.
Still some could be light, even very light, 
provided they are, also, very weakly coupled -- and therefore neutral.

\vspace{.5mm}

We discussed, long ago, the possible existence of such a new gauge boson called $\,U$,
exploring in particular
limits on its production and decay (depending on its mass)
into $\,e^+e^-$ or $\,\nu \bar\nu\,$ pairs ...\cite{fayet:1980rr}.
Such a particle originated from supersymmetric extensions of the Standard Model, 
which require two electroweak doublet Higgs superfields
\cite{susy},
offering the possibility, in non-minimal versions\,\footnote{The
$\,\mu\,H_1H_2\,$ superpotential term of the MSSM was then replaced, as in \cite{nmssm}, by a trilinear coupling $\,\lambda \,H_1H_2\,N\,$ with an extra singlet chiral superfield
$N$, which now transforms
under the extra-$U(1)$ symmetry.},
of ``rotating'' independently the two doublets, i.e. of gauging an extra-$U(1)$ symmetry.  
The fact that the effects of such a gauge boson did not show up experimentally (and a possible connection with gravity through the massive gravitino)
led us to consider that it could be both light, 
and very weakly coupled.
In any case, independently of its possible origin,
the phenomenology of a light neutral \linebreak
spin-1 $U$ boson 
turns out to be quite rich.
It could be produced in $ \,q \bar q \,$ or $\,e^+ e^-\,$ annihilations through processes like
\be
\label{prodU}
\psi \to \gamma\ U\,, \ \ \Upsilon \to \gamma\ U\,, \ \ K^+\to\ \pi^+\,U\,,\ \ \hbox{or}\ \
e^+e^-\,\to \gamma\ U\,,
\ee
including even positronium decay, should the $U$ be lighter than 1 MeV \cite{fayet:1980rr,fayetmez}.
It could also lead to interesting effects in neutral-current phenomenology, 
including neutrino scatterings, anomalous magnetic moments of leptons, 
and parity-violation in atomic physics ...  \cite{fayet:1980rr,pvat}.

\vspace{.5mm}

If the extra-$U(1)$ gauge coupling \,($g"$)\, is very small it looks like the $U$ 
will be very weakly coupled.
Still the rates for producing a light $\,U$, although seemingly $\propto g"^2$,
would not necessarily be so small if this particle has axial couplings.
In the low mass and low coupling regime a light \hbox{spin-1} $U$ boson would then be produced very much like a spin-0 axion,
proportionally to $\,g"^2/m_U^{\,2}$
(just like a light spin-$\frac{3}{2}$ gravitino, although gravitationally coupled, 
would be produced and interact very much like a spin-$\frac{1}{2}$ goldstino, 
proportionally to $\kappa^2/m_{3/2}^{\,2}$ \cite{susy}).
Searches for $U$ bosons, as in the hadronic decays (\ref{prodU}),
decaying into unobserved $\nu\bar\nu$ or light dark matter particle pairs,
then require the  extra-$U(1)$ to be broken at a scale higher 
than electroweak;
and possibly at a large scale if an extra singlet acquires a large v.e.v., 
according to a mechanism already pointed out in \cite{fayet:1980rr}, 
which also applies to spin-0 axions as well.
(Furthermore, if the $U$  were extremely light or even massless, with extremely small couplings, 
a new force could lead to apparent deviations from the $1/r^2$ law of gravity,
and violations of the equivalence principle\,\cite{equiv}.)

\vspace{.5mm}

Such a $U$ boson may also play a role in the annihilation of dark matter particles.
While weakly-interacting massive particles must in general be rather heavy,
one may now consider 
{\it light} dark matter (LDM) particles, 
by introducing new efficient mechanisms responsible for their annihilations.
The $U$ boson, very weakly coupled but still leading to relatively ``large'' annihilation cross sections, may then lead
to the right relic abundance ($\Omega_{\rm dm}\simeq 22\,\%$) for the non-baryonic dark matter of the Universe;
exchanges of charged heavy (e.g. mirror) fermions could play a role too,
for \hbox{spin-0} LDM particles \cite{boehmfayet}.
$\,U$-induced annihilations also allow
for a $P$-wave cross section, as useful to avoid a potential danger of excessive $\gamma$-ray production \cite{bes} (depending, however, on how this production occurs and is estimated).

\vspace{.5mm}

The subsequent observation by INTEGRAL/SPI of a bright 511 keV
$\gamma$-ray line from the galactic bulge \cite{integral}
could then be viewed as a sign of the annihilations of positrons
from light dark matter annihilations \cite{boehm511}.
These particles, that could explain both the {\it \,non-baryonic dark matter\,} 
and {\it \,the 511 keV line}, may have spin $\frac{1}{2}$ instead of spin 0 \cite{fermion}.
The new annihilation processes mediated by $U$ exchanges appear as {\it stronger than weak interactions}, at lower energies, while becoming {\it weaker than weak} (and difficult to detect)
 at higher energies.
The mass of the  $\,U$ boson and its couplings 
to leptons and quarks are already strongly constrained,
independently of dark matter.
There are also several constraints from cosmology and astrophysics 
involving the characteristics of the LDM particles $\chi$,
should the $U$ be responsible for their annihilations.

\vspace{2mm}
{\bf Dark Matter requirements:}\ \ 
i) the total LDM annihilation cross section 
at freeze out should be $\simeq 4$ or 5 pb, 
to get the right relic abundance; more precisely \cite{fermion}:
\be
\label{sigmaF}
<\!\sigma_{\rm ann}\,v_{\rm rel}/c\!>_F\, \simeq 4\ \hbox{to 5 pb}
\left\{\!\ba{l}
\small \times \ 2\ \  \hbox{if LDM not self-conjugate,}
\vspace{2mm}\\
\small \times \ \frac{1}{2} \  \hbox{if} \ S\ \hbox{instead of $P$-wave ann.}
\ea\right.
\ee

\vspace{.5mm}

ii) constraints from the intensity of the 511 keV $\gamma$-ray line from the galactic bulge involve the annihilation cross section for $\,\chi\,\chi\to e^+e^-\,$
at low halo velocities, and depend on whether it is $S$-wave or $P$-wave-dominated (with $\,\sigma_{\rm ann}\,v_{\rm rel}\,\propto 1\,$ or $\,v^2$, \,respectively). 
They are also sensitive to the shape of the
dark matter profiles adopted within the bulge \cite{asc,rasera}.

\vspace{1mm}
A $S$-wave cross section, such that 
$<\!\sigma_{\chi\chi\to e^+e^-}v_{\rm rel}/c>_{\rm halo}$ $\approx\, <\!\sigma_{\chi\chi\to e^+e^-}\,v_{\rm rel}/c>_F \ \approx\,$ 
1 to a few pb
\footnote{It should be $\simeq 2$ pb (doubled in the non-self-conjugate case),
times the branching ratio $B_{\rm ann}^{ee}$ for producing $e^+e^-$ 
in LDM annihilations. 
This $B_{\rm ann}^{ee}$ could be $\simeq 40$\% if all decay channels into
$\,e^+e^-$ or $\,\nu\bar\nu\,$ pairs contribute equally;
it could also approach 1, as $\,\nu\bar\nu$ modes may well be suppressed.},
\,would then necessitate a (relatively) heavier LDM particle, say
 $\gsim 30$ MeV (as the LDM number density scales as $ 1/m_\chi$ and the 511 keV 
emissivity as $1/m_\chi^2$), which is probably excluded
as we shall see.
A $P$-wave cross-section, for which $\,<\!\sigma\,v_{\rm rel}\!>_{\rm halo}\,$ would be much smaller, 
would require, to get the observed 511 keV signal, a much lighter LDM particle 
( $\simeq\frac{1}{2}$ to typically a few MeV), 
with a rather peaked halo profile~\cite{rasera} (cf. Fig.~7)\,\footnote{The profile should be sufficiently steep near the Galactic Center, e.g. a ``Moore-type'' distribution 
with $\rho \approx r^{-\gamma}$ not so far from $r^{-1.5}$, \,near the
Galactic Center.}, or a more clumpy one (in which case the mass could be higher).
Intermediate situations are also possible for a wide range of LDM masses, 
with a cross-section (\ref{sigmaF}) \,\hbox{$P$-wave} dominated at freeze-out, 
later becoming smaller and ultimately $S$-wave dominated (or $\,S+P$-wave) 
for low-velocity halo particles \cite{asc,rasera}
\footnote{In particular, we may have $S$-wave-dominated halo annihilations,
with typical $m_\chi\approx 3$ to \,30\, MeV, and
a cross section (scaling like $\,m_\chi^{\,2}$) which depends on the dark matter profile:
\vspace{-1.5mm}
$$
\vspace{-1mm}
\ \ \ \ \ \ (\sigma_{\chi\chi\to e^+e^-}v_{\rm rel})_{\rm halo} \approx(.2\ \hbox{fb}\ \, \hbox{to}\ \, .2\ \hbox{pb})\ 
\hbox{\small $\displaystyle \left(\,m_\chi/(10\ \hbox{MeV})\,\right)^2$},\!\!\!
$$
as can be seen from Fig.~7 of \cite{rasera}, small compared to the cross section at
freeze-out,
to be provided by the $P$-wave term.}.

\vspace{1mm}

Other constraints (iii) require that the LDM mass $m_\chi$ 
be sufficiently small (say $\lsim 30$ MeV),
to avoid excessive $\gamma$-rays
from inner-bremsstrahlung, bremsstrahlung, and in-flight annihilations~\cite{fermion,beacom}. 
Constraints (iv) from core-collapse supernovae require them to be 
$\gsim 10$ MeV at least, if they have relatively ``strong'' interactions with neutrinos,
as they do with electrons~\cite{fhs}. No further constraints are obtained from the soft $\gamma$-ray extragalactic background that could be gener-
\linebreak ated by the cumulated effects of LDM annihilations
\cite{rasera}.

\vspace{3mm}

{\bf \boldmath $e^+e^-\to\gamma\,U\,$ and LDM annihilations:}\ 
$\,U$ bosons may be directly produced in an accelerator experiment,
through $\,e^+e^- \to \gamma\ U$
\cite{fayetmez,boehmfayet,boro}.
The relevant parameters are the mass $m_U$ and the couplings 
$\,f_{eV}$ and $\,f_{eA}$ to the electron, expressed in terms of chiral couplings as $\,(f_{eL}+f_{eR})/2\,$ and $\,(f_{eL}- f_{eR})/2$.
The detectability of the process, of order 
\be
2\ \ \frac{f_{eV}^2+f_{e\,A}^2}{e^2}\ \ =\ \ \frac{f_{eL}^2+f_{eR}^2}{e^2}\ \ \ \ \ \ \ 
(\ =\ 2\ \,\frac{f_e^2}{e^2}\ )\ \ ,
\ee
 as compared to $\,e^+e^-\to \gamma\gamma$, \,depends essentially on the values of the $U$
couplings to the electron.
At energy $\,E\,$ large compared to $m_U$, \,one has
\be
d\sigma\ (e^+e^-\to \gamma\,U)\ \simeq\ \frac{f_{eL}^{\,2}+f_{eR}^{\,2}}{e^2}\ \ 
d\sigma\ (e^+e^-\to \gamma\,\gamma)\ \ ,
\ee
with
$\,
\frac{d\sigma}{d\cos\theta }\ (e^+e^-\to \gamma\,\gamma)\, \simeq \, \frac{4\pi\,\alpha^2}{s}\ (\,\frac{1}{\sin^2\theta}-\frac{1}{2}\,)\,,
$
\,so that 
\,($\theta$ being now in the $\,[0,\pi]\,$ instead of $\,[0,\pi/2]\,$ interval ):
\be
\frac{d\sigma}{d\cos\theta}\,(e^+e^-\!\to \gamma\,U)\ \simeq \ \frac{\alpha\ 
(f_{eL}^{\,2}+f_{eR}^{\,2})/2}{s}\ \,(\,\frac{1}{\sin^2\theta}\,-\,\frac{1}{2}\,)\,.
\ee

\vspace{1mm}
The $U$ boson can then decay into $e^+e^-$, or an invisible $\nu\bar\nu$ or LDM particle pair (the latter being favored for $m_U>2\,m_\chi$)
\footnote{$U\to\,e^+e^-\,$ may represent only $\approx$ 40 \% of the decays, 
if all $e^+e^-$ and $\nu\bar\nu$ channels contribute equally,
the $\chi\chi$ mode being forbidden.
$B_{U\to e^+e^-}$ could also be very close to 0 if $m_U>2m_\chi$, 
as the $U$ is expected 
to be more strongly coupled to LDM than to ordinary particles.
It could approach 1 if $m_U<2\,m_\chi$, with the $U$ coupling
much less to neutrinos than to electrons.}.
The possibility of detecting them at 
current $B$-factories or at the $\phi$ factory DA$\Phi$NE, which could be sensitive to couplings $f_{eR}$ larger than $10^{-4} - 10^{-3}$  (DA$\Phi$NE)
down to $3\ 10^{-5} \,-\, 3\ 10^{-4}$ ($B$-factories), has been considered recently (the first numbers correspond to 
100\,\% invisible decay modes, the last to 100\,\% decays into $e^+e^-$) \cite{boro}. This analysis, however, ignored that coupling 
a new gauge boson to $e_R$ but not to $\nu_L$ or $e_L$ would necessitate a coupling 
to a Higgs doublet, usually inducing {\it mixings} between electroweak and extra-$U(1)$ 
gauge bosons, and requiring couplings to quarks as well.
It also disregarded very stringent constraints associated with {\it axial couplings} 
of the $U$. We would like to discuss here which values of the couplings to electrons are actually possible, and under which circumstances.

\vspace{.3mm}
Annihilation cross sections of LDM particles into $e^+e^-$ depend on the product $\,c_\chi f_e$
\,($c_\chi$ denoting the $U$ coupling to the LDM particle $\chi$), as well as on $m_U$ and $\,m_\chi$, and more precisely on
$\,\frac{c_\chi\,f_{e}}{m_U^{\,2}-4\,m_\chi^{\,2}}\  m_\chi\,$.
To get an annihilation cross section into $e^+e^-$ of the order of 4 to 5 pb, times the branching fraction $B_{\rm ann}^{ee}$, 
as follows from (\ref{sigmaF}), we need (cf. eq.~(16) of \cite{fermion}):
\be
\label{sizecf}
|c_\chi|\ (f_{eV}^{\,2}+f_{eA}^{\,2})^\frac{1}{2}\ \simeq 10^{-6}\ 
\frac{|m_U^{\,2}-4\,m_\chi^{\,2}|}{m_\chi\ (1.8\ {\rm MeV})}\ \ 
\left(B_{\rm ann}^{ee}\right)^{\frac{1}{2}}.
\ee
For $m_U= 10$  MeV 
and $ m_\chi = 4 $ MeV (cf.~\cite{boehmfayet}), or 6 MeV,
this would give 
$\ 
|c_\chi\,f_e|\, \simeq\,5\ 10^{-6}\,$ 
(or $\,\simeq 3\ 10^{-6}$ if only 40\% of annihilations led to $e^+e^-$).
For a heavier $U$ we could get larger couplings, e.g. up to
$|c_\chi\,f_e|\simeq  10^{-2}/$ $(2\ m_\chi(\hbox{MeV)})\, $ for a 100 MeV $U$.

\vspace{.3mm}

Discussing, however, limitations on $c_\chi f_e$ 
does not help so much as we are primarily interested in $f_e$.
Dividing it by 10 while multiplying $c_\chi$ by 10\,
leaves unchanged the annihilation cross section 
at freeze out (and nowadays in the halo);
but it has a crucial effect on the detectability of the $U$ by dividing 
its production cross section by 100.
In fact,
{\it dark matter considerations only play a secondary role},
once we have checked that suitable cross sections can indeed be obtained,
with an appropriate $c_\chi \lsim 1$ or in any case $\sqrt{4\,\pi}\ $ if we would like 
the theory to remain perturbative.
Still $m_U$ should better not be too large as compared to $2\,m_\chi$, otherwise the 
$U$ couplings to ordinary particles would tend to be too large if $c_\chi$ is to remain perturbative.
Demanding $\,c_\chi<\sqrt{4\,\pi}\,$ would imply from (\ref{sizecf}) 

\vspace{-4mm}
\be
\label{sizef}
f_e=(f_{eV}^{\,2}+f_{eA}^{\,2})^\frac{1}{2}\ \gsim \ 3\ 10^{-7}\ 
\frac{|m_U^{\,2}-4\,m_\chi^{\,2}|}{m_\chi\ (2\ {\rm MeV})}\ \ 
\left(B_{\rm ann}^{ee}\right)^{\frac{1}{2}}.
\ee
For $m_U=10$ MeV and  $m_\chi=4$ (or 6)\, MeV,
it should then verify $\,f_e\gsim 10^{-6}$, with
$B_{\rm ann}^{ee}\simeq 1$.
For $m_U=100$ MeV with $m_\chi=5$ \,(resp. 15) MeV, 
$\,f_e \gsim 3\ 10^{-4}$ (resp. $10^{-4}$);
for $m_U=300$ MeV with $m_\chi=15$ MeV, 
$\,f_e \gsim 10^{-3}$.

\vspace{1.5mm}

{\bf Constraints from \boldmath $\,g_e\!-2\,$:}\ 
for a {\it vector coupling} 
to the electron, the extra contribution to $\,a_e\!=(g_e\!-2)/2$ is given by
\be
\delta a_e\simeq 
\frac{f_{e V}^{\, 2}}{4\ \pi^2}\int_0^1\!\frac{m_e^{\,2}\ x^2\,(1-x)\ dx}
{m_e^{\,2}\,x^2+m_{U}^{\,2}(1-x)}
\simeq \frac{f_{eV}^{\ 2}}{12\ \pi^2}\ \frac{m_e^{\,2}}{m_U^{\,2}}\ F(\hbox{\small$\displaystyle\frac{m_U}{m_e}$}),
\ee
with $F\simeq  \pi/\sqrt 3 -3/2 \simeq .31;\, .54,\, .81,\, .92$, or $\simeq 1$ for
$m_U=$ $m_e;\, 2m_e,\, 5m_e, \, 10m_e$, or large.
Satisfying
$\,\delta a_e\simeq (1.24\pm .95)\, 10^{-11}$ \cite{kinoshita},
i.e. 
$
\,-\,10^{-11} \lsim\delta a_e \lsim  3\, 10^{-11}$,
\,requires
\be
\label{limfve}
|f_{eV}|\ \lsim \ 1.3\ 10^{-4}\ m_U(\hbox{MeV})\ \ ,
\ee as soon as $m_U\gsim 2$ MeV (the limit being slightly weaker otherwise).
If there is also an {\it axial coupling} one gets \cite{fayet:1980rr,pvat}
\be
\delta a_e\simeq \frac{f_{eV}^{\ 2}-5\,f_{eA}^{\ 2}} {12\ \pi^2}\ \frac{m_e^{\,2}}{m_U^{\,2}}
\simeq \frac{3\,f_{eL}\,f_{eR}-f_{eL}^{\ 2}-f_{eR}^{\ 2}}{12\ \pi^2}\ \frac{m_e^{\,2}}{m_U^{\,2}}
\,,
\ee
which implies, roughly, for $m_U\gsim$ a few MeV,
\be
-5\ 10^{-9}\, m_U(\hbox{MeV})^{\,2} \!\lsim
f_{eV}^{\ 2}-5\,f_{eA}^{\ 2}
\lsim
1.5\ 10^{-8}\, m_U(\hbox{MeV})^{\,2}.
\ee
While no limit can in general be obtained due to possible compensations, one gets
$|f_{eA}| \lsim  3\ 10^{-5}\  m_U(\hbox{MeV})$ for an axial coupling, 
and
$|f_{eR}|\lsim 7\ 10^{-5}\  m_U(\hbox{MeV})$ for a chiral one
\footnote{Such limits could be alleviated in the presence of extra contributions 
to $a_e$ and $a_\mu$, as from heavy (e.g. mirror) fermions
in the case of spin-0 LDM particles \cite{boehmfayet}.
Conversely, $U$ exchanges could help making acceptable such heavy fermion contributions 
to $g-2$, \,if present.
}.
\vspace{1mm}

{\bf Constraints from \boldmath $\ g_\mu-2\,$:}\ \
for a $U$ with  a {\it vector coupling} 
to the muon, one has, similarly,
\be
\delta a_\mu\,\simeq \frac{f_{\mu V}^{\, 2}}{4\ \pi^2}\,\int_0^1\ \frac{m_\mu^{\,2}\ x^2\,(1-x)\ dx}
{m_\mu^{\,2}\ x^2+m_{U}^{\,2}\,(1-x)}\ \simeq
\frac{f_{\mu V}^{\, 2}}{8\ \pi^2} \ G(\hbox{$\displaystyle\frac{m_U}{m_\mu}$})\ ,
\ee
with $\,G\simeq1, \ .77,\ .57,\ .38,$ or $2 \pi/3 \sqrt 3 -1\simeq .21$, \,for $\,m_U$ small,
or $m_\mu/10,$ $m_\mu/4,\ m_\mu/2$, or $m_\mu$.
Satisfying
$\,
\delta a_\mu\simeq (2\pm 2)\ 10^{-9}\,$,
i.e., to be conservative,
$-2\ 10^{-9} \lsim \delta a_\mu\lsim 6\ 10^{-9}$,
leads in this mass range to
\be
|f_{\mu V}|\ \lsim \ (.7\ \ \hbox{up to}\ \ 1.5)\ \ 10^{-3}\ \ .
\ee
In the natural case of a universal coupling to charged leptons, 
this limit is more constraining than (\ref{limfve}),
\,for $m_U \gsim \,6$  MeV.

\vspace{2mm}

If the coupling has also an {\it axial part}, one has
\be
\delta a_\mu\ \,\simeq\, \ \frac{f_{\mu V}^{\ 2}}{8\ \pi^2}\ G(\hbox{$\displaystyle\frac{m_U}{m_\mu}$})\ -\
\frac{f_{\mu A}^{\ 2}} {4\ \pi^2}\ \frac{m_\mu^{\,2}}{m_U^{\,2}}\ 
H(\hbox{$\displaystyle\frac{m_U}{m_\mu}$})
\ \ ,
\ee
the second term being associated with an ``axionlike'' behavior of the exchanged $U$ boson, when this one is light
\cite{fayet:1980rr,pvat}.
More precisely \footnote{
\vspace{.5mm}
One has:\ \
$\displaystyle
H=\int_0^1\frac{2x^3+(x-x^2)(4-x)\,m_U^{\,2}/m_\mu^{\,2}}{x^2+(1-x)\,m_U^{\,2}/m_\mu^{\,2}}\ \ dx
\ \ .
$
},
one has $ H \simeq 1,\ 1.18,\ \pi/\sqrt 3 - 1/2 \simeq 1.31,\,$ or
$\,\to 5/3$, \,for
$m_U$ small, $m_\mu/2,\ m_\mu$, or large, respectively.
A purely axial coupling would then have to verify
\be
|f_{\mu A}|\,\lsim\ 3\ 10^{-6}\ m_U(\hbox{MeV})\ \ ,
\ee
also expressed as 
$\,
\frac{f_{\mu A}^2}{m_U^{\ 2}}\lsim G_F\,.$
\,Similarly, for a $U$ sufficiently light compared to $m_\mu$, we would get
$\,
|f_{\mu R}|\lsim 6\ 10^{-6}$ $m_U(\hbox{MeV})\,$.
\,These limits are more restrictive than those from $\,g_e-2$, \,by about an order of magnitude.

\vspace{2mm}

Taking both $\,g_e-2\,$ and $\,g_\mu-2\,$ into consideration and assuming lepton universality,
we get
\be
\label{fl}
\ba{l}
|f_{l V}|\lsim  
\left\{ \ \ba{lc}
1.3\ 10^{-4}\ m_U(\hbox{MeV})&\!\!\!\!\!\hbox{\small(2 MeV$< m_U \lsim 6\ \hbox{MeV})$}\,,\vspace{1mm}\\
\ \ 7\ 10^{-4} \ \ \hbox{up to}\ \ 1.5 \ 10^{-3}& \hbox{\small$(m_U < 100 \ \hbox{MeV})$}\,,
\ea \right.
\vspace{3mm} \\
\hbox{or}\ \ \ \ |f_{l A}|\ \lsim\ \ 3\ 10^{-6}\ \,m_U(\hbox{MeV})\ \ ,
\vspace{-2mm}
\ea
\ee
or 
$\,|f_{l R}|\lsim 6\ 10^{-6}\ m_U(\hbox{MeV})\,$ in the chiral case.
This in general decreases, especially for axial couplings, the
maximum production cross section, compared to what could be inferred from $\,g_e-2\,$
only.

\vspace{2mm}

{\bf Restrictions \,from \,quark \,couplings:}\ 
the easiest way through which a $U$ boson could manifest, and in general 
be quickly excluded, would be through {\it flavor-changing neutral current} processes.
Fortunately in the simplest cases its couplings to quarks are found
to be flavor-conserving, as a consequence of the extra-$U(1)$ gauge symmetry
of the (trilinear) Yukawa interactions responsible for quark and lepton masses,
which naturally avoids prohibitive FCNC processes \cite{equiv,U}.

\vspace{1mm}
{\it Searches for axionlike particles} in the decays $\,\psi\to\gamma\,U$, 
$\,\Upsilon\to\gamma\,U$, with the $U$ decaying into unobserved LDM 
or $\nu\bar\nu$ pairs, strongly constrains possible axial couplings to heavy quarks:
\be
|f_{cA}|\ \lsim 1.5 \ 10^{-6} \ m_U(\hbox{MeV})\ ,\ \ 
|f_{bA}|\ \lsim .8 \ 10^{-6} \ m_U(\hbox{MeV})\ \ ,
\ee
already implying that the extra-$U(1)$ symmetry should be broken at least somewhat above the electroweak scale \cite{fayet:1980rr,fermion}; one may also get, by searching for the decay $\,K^+\to\pi^+ U$,
\be
|f_{sA}|\ \lsim 2 \ 10^{-7} \ m_U(\hbox{MeV})\ \ .
\ee

\vspace{-.4mm}

Experiments looking for {\it parity-violation effects in atomic physics} 
constrain the product of the axial coupling of the $U$ to the electron, times its
(average) vector coupling to a quark, to be very small, typically
$\,\frac{|f_{eA}\,f_{qV}|}{m_U^{\ 2}}\lsim\,10^{-3}\ G_F$, \,or more precisely \cite{pvat}:
\be
-1.5\ 10^{-14}\ m_U(\hbox{MeV})^2 \lsim f_{eA}\,f_{qV} \lsim .6\ 10^{-14}m_U(\hbox{MeV})^2.
\ee
These limits, valid in the local limit approximation for $m_U \geq 100$ MeV, should be multiplied by a corrective factor $K^{-1}(m_U) \geq 1$,
\,which is about 2 for  $m_U$ of a few MeV's.

\vspace{1mm}
Axial couplings to the electron that would approach a few times $10^{-5}\ m_U$(MeV), 
as considered previously (only from $g_e-2$), would require the effective vector coupling 
to quarks to be extremely small, 
$
|f_{qV}|\lsim \,10^{-9} \ m_U(\hbox{MeV})\,
$.
More conservatively, having $\,|f_{eA}| \gsim  10^{-6}\ m_U$(MeV)
(as considered most of the time in \cite{boro}) 
would require $|f_{qV}| \lsim$ a few $10^{-8} \ m_U$(MeV), 
still very restrictive.

\vspace{1mm}

A simple way to satisfy automatically such stringent limits 
on axial couplings would be to consider situations,
natural in a number of models,
in which the $U$ couples to leptons and quarks {\it in a purely vectorial way} \cite{equiv,U}.

\vspace{2mm}

{\bf Other constraints \,on \,lepton \,couplings:}\ \,
but maybe the $U$ does not couple to quarks at all\,?
As quarks and leptons usually acquire their masses through trilinear Yukawa couplings 
to the same Higgs doublet (or doublet pair, in a supersymmetric theory), 
demanding that the extra $U(1)$ does not act on quarks 
implies that it does not act on Higgs doublets either,
leading to a $\,U$ current proportional to the leptonic current
(or to $\,L_e$, 
\,or $\,L_e -L_\mu$, \,or $\,L_e-L_\tau$, \,...), \,plus an additional dark matter contribution.

\vspace{1mm}
But we still have to take into account  
another stringent constraint
in the purely leptonic sector, namely, from low-$|q^2|$ \,$\nu$-$e$  scattering,
$\,\frac{f_\nu\ f_{e}}{m_U^{\,2}} \ \lsim \ G_F\,$,
\,for $m_U$ larger than a few MeV's \cite{boehmfayet}.
If the $U$ couplings to $e$'s and $\nu$'s, $f_e$ and $f_\nu$, are taken to be equal
(or of the same order)  they would have to verify
\be
\label{fe}
f_{e}\,\lsim\, 3\ 10^{-6}\ m_U(\hbox{MeV})\ \ ,
\ee
which is about $\,3\ 10^{-5}\,$ at 10 MeV, 
reducing further (compared to the $\approx 10^{-4}\ m_U$(MeV) of (\ref{limfve}) 
or $\approx 10^{-3}$ of (\ref{fl})) the hopes of detecting a light $U$ in $e^+e^-$ annihilations);  \,up to $\,\approx 10^{-3}$ at 300 MeV.
This upper limit (\ref{fe}) is still larger than the lower one (\ref{sizef})
from the annihilation cross section, using the requirement that the  coupling to $\chi$ remains perturbative
(unless $\,m_U$ is taken too large as compared to $2\,m_\chi$).

\vspace{2mm}
{\bf \boldmath $Z$-\boldmath$U$ \,mixing \,effects:}\
one may also satisfy the above leptonic constraint 
while allowing for larger couplings to electrons 
by having {\it very small or vanishing couplings to $\,\nu$'s}.
This requires taking into account, to our advantage, mixing effects
between the $Z$ and $U$, as the extra $U(1)$ should now act on Higgs fields as well.
The Higgs v.e.v.'s then in general induce a mixing 
between the standard $Z$ and the extra-$U(1)$ gauge field. This mixing remains tiny as the extra-$U(1)$ coupling is very small, and does not significantly affect the $Z$, \,while
the $U$ current picks up an extra part proportional to  
$\,J^\mu_{\,3}-\sin^2\theta \,J^\mu_{\rm \, em}$~\cite{equiv,U} .
The vector part in the $U$ current then normally appears
as a combination of the $B$, $L$ and electromagnetic currents; the axial part 
may well be completely absent, depending on the theory considered (there is of course, in addition, an extra LDM part).
This provides a favorable situation, in view of having
 ``large'' (vectorial) couplings to electrons, bounded by (\ref{fl}).

\vspace{.5mm}
If, however, the $U$ couplings to electrons and neutrinos turn out to be
similar, they should again verify as in (\ref{fe})
$\,f_{e}\lsim 3\ 10^{-6}\ m_U(\hbox{MeV})\,$,
\,much more constraining than the $\,\approx 10^{-4}\ m_U(\hbox{MeV})$ of (\ref{limfve}) from $\,g_e-2$. 
{\it Still a 100 MeV (300 MeV) $U$ would allow for 
a coupling to the electron of up to $\,\approx3\ 10^{-4}$ (resp. $10^{-3}$), that could be detectable},
especially if the $U$ decays invisibly into $\chi\chi$ pairs.

\vspace{2mm}
{\bf In summary,} \ constraints which do not involve dark matter directly,
as from an axionlike behavior of a $U$ boson, or atomic-physics parity-violation, 
as well as $Z$-$U$ mixing effects, 
cannot be ignored.
A favorable situation
is obtained when couplings to quarks and leptons are {\it vectorial},
with \,-- {\it thanks to $Z$-$U$ mixing effects} --\, a much smaller coupling to neutrinos
than to electrons 
\cite{U} (as also useful to obey the supernovae constraint on lighter 
dark matter particles \cite{fhs}).
The $g-2$ constraints (\ref{fl}), in the absence of cancellations, allow for 
a vectorial coupling to charged leptons of up to $\,\approx (.7$ to $1.5)\ 10^{-3}\,$ 
for $\,m_U<m_\mu$ (from $\,g_\mu-2\,$ assuming lepton universality).
Having $f_e^{\,2} \lsim 10^{-5}\,e^2$  makes the detection of $U$ production in $e^+e^-$ colliders 
difficult.
The prospects for actually producing and detecting such very weakly coupled $U$ bosons in $e^+e^-\to\,\gamma\,U$, as well as in other reactions,
appear as challenging. Still efforts should be pursued in this direction.

\end{document}